\documentclass[preprint]{elsarticle}

\usepackage{amssymb, latexsym, amsthm, amsmath, lineno, epsfig, mathtools, hyperref, todonotes, booktabs, cite, graphicx}
\usepackage[singlelinecheck=false]{caption}
\modulolinenumbers[1]

\journal{Theoretical Population Biology}

\newcommand{\bs}[1]{\ensuremath{\boldsymbol{#1}}}

\newcommand\given{{\,|\,}}

\newcommand\eg{{\it e.g.,}}

\newcommand\ie{{\it i.e.,}}

\newcommand\x[1]{\ensuremath{x_{#1}}}
\newcommand\y{\ensuremath{y}}

\newcommand\fv[1]{\ensuremath{\mathbf{f}_{#1}}}
\newcommand\bv[1]{\ensuremath{\mathbf{b}_{#1}}}

\begin{document}

\author[address1,address2]{Claus Vogl}
\cortext[correspondingauthor]{Corresponding author}
\ead{claus.vogl@vetmeduni.ac.at}
\author[address1,address2]{Sandra Peer}
\ead{sandra.peer@tuwien.ac.at}
\author[address3]{Lynette Caitlin Mikula\corref{correspondingauthor}}
\ead{lcm29@st-andrews.ac.uk}

\address[address1]{Department of Biomedical Sciences, Vetmeduni Vienna, Veterin\"arplatz 1, A-1210 Wien, Austria}
\address[address2]{Vienna Graduate School of Population Genetics, A-1210 Wien, Austria}
\address[address3]{Centre for Biological Diversity, School of Biology, University of St. Andrews, St Andrews KY16 9TH, UK}

\begin{frontmatter}


\title{Forward-backward algorithms with a biallelic mutation-drift model: Orthogonal polynomials, and a coalescent/urn-model based approach}

\begin{abstract}
 Inference of the marginal likelihood of sample allele configurations using backward algorithms yields identical results with the Kingman coalescent, the Moran model, and the diffusion model (up to a scaling of time). For inference of probabilities of ancestral population allele frequencies at any given point in the past - either of discrete ancestral allele configurations as in the coalescent, or of ancestral allele proportions as in the backward diffusion - backward approaches need to be combined with corresponding forward ones. This is done in so-called forward-backward algorithms. In this article, we utilize orthogonal polynomials in forward-backward algorithms. They enable efficient calculation of past allele configurations of an extant sample and probabilities of ancestral population allele frequencies in equilibrium and in non-equilibrium. We show that the genealogy of a sample is fully described by the backward polynomial expansion of the marginal likelihood of its allele configuration. 
\end{abstract}


\end{frontmatter}


\section{Introduction}

In population genetics, extant (present-time) data are usually used for inference of past population genetic processes. The coalescent \citep{King82} is a stochastic process that describes the genealogy of a sample from a single locus back to the last common ancestor. It allows for convenient simulations of genealogical trees, conditional on the current sample size and either the current (effective) population size or past (effective) population sizes. Along the branches of this tree, mutations may introduce new allelic states. The coalescent has become a pillar of theoretical and empirical population genetics (see \citep[\eg][]{Hein05,Wake09} and \eg\ the software {\it ms} \citep{Huds90}). 

While the extant sample is assumed given, many intermediate configurations are possible until the last common ancestor of the sample is reached. Mutations increase this complexity further. Coalescent simulations have been used for inferring parameters of relatively complex population genetic models assuming high-throughput population genetic data. The combinatorial complexity of the coalescent with mutation means that sufficient statistics are often not available. Summary statistics and Approximate Bayesian Computation (ABC) \citep[\eg][]{Beau02,Esto10,Frai17} are then employed for inference of population genetic parameters. Summary statistics are generally not sufficient and ABC is so computationally demanding that often only subsets of the parameter space can be investigated. Inference therefore becomes approximate. 

The infinite sites model \citep{Kimu71} is another classic population genetic model; one of its assumptions is complete linkage disequilibrium, \ie\  no recombination \citep[\eg][]{Watt75}. Using a forward algorithm (a dynamic programming method) with the underlying logic of an urn model, \citet{Wu10} showed that inference under the infinite sites model is efficient in analysis of data sets from panmictic or subdivided populations in equilibrium. Forward in time, the sample sizes leading to the extant sample size are random variables. Therefore, the forward approach cannot be extended to account for deviations from equilibrium, \eg\ via changing population sizes. \citet{Fais15} introduced a corresponding backward algorithm; since it proceeds backward in time conditional on the extant sample, modelling changing population sizes is possible. 

Note that data from autosomes of some model species are further from complete linkage disequilibrium, which is assumed by the infinite sites model, than from linkage equilibrium: In flies of the genus {\it Drosophila,} deviations from linkage equilibrium are barely noticeable in genomic regions of moderate to high recombination rates \citep{Pars10} and weak enough to be negligible even within short introns \citep{Clem12a}. In great apes, recombination rates are considerably more variable across the genome: while on average neither linkage equilibrium nor complete disequilibrium can be assumed, either assumption may hold approximately in certain genomic regions (see Supplementary Material, Figure S3 in \citep{Myers05}). 

In the protein coding genes of eukaryotes, the expected heterozygosity (which is roughly equal to the scaled mutation rate) is approximately $\leq 10^{-2}$ or smaller \citep{Lynch06}. This makes it unlikely to find more than one allele segregating in a small to moderately sized sample. Thus bi-allelic mutation models suffice to capture the true dynamics of a population. In genomic regions where the recombination rate is much greater than the mutation rate, polymorphic sites can additionally be considered independent: The distance between polymorphic loci in small to moderately sized samples is great enough for recombination to break their association. In other words, neighboring polymorphic sites have different genealogies due to recombination. Data from such populations can be represented in a site frequency spectrum (SFS) without loss of information (and data from multiple populations in a joint site frequency spectrum (jSFS)). The SFS records the frequency of the focal allele in a sample. A bi-allelic mutation model that can be represented in this way can also be re-parametrized to a parent-inde\-pen\-dent mutation model; this means that both mutation and coalescent events are uninformative regarding the immediately ancestral allelic state. 

In this article, we assume data that can be represented as site frequency spectra without loss of information. We further assume that data are generated by a haploid bi-allelic Moran model of population size $N$ or by the corresponding diffusion model. (Note that the Wright-Fisher model has the same diffusion limit as the Moran model.) In Sec.~(\ref{section:recap_Moran}), we first review the discrete time and discrete space decoupled mutation-drift Moran model and the derivation of its corresponding diffusion equation. Then we describe two known forward-backward algorithms previously applied by \citet{Berg17}. Such algorithms enable inference of probabilities of ancestral population allele frequencies at any time in the past in addition to the probabilities of extant sample configurations. 

The first of these algorithms is for the discrete case: Essentially, the algorithm of \citet{RabinerJuang86} can be directly applied \citep{Berg17}. Our representation of this algorithm is based on matrix multiplications and the population size is assumed constant over time. Changes in mutation-drift parameters can however be immediately incorporated and changes in the effective population size can be indirectly modeled through rescaling of time. 

The second of these algorithm is for the continuous case: An analytical form of the transition density function is found by decomposition of the diffusion generator into eigenvectors of orthogonal polynomials (more precisely, modified Jacobi polynomials) and their corresponding eigenvectors \citep{Song12,Berg17}. The prior distributions of both the ancestral population allele proportions and the probabilities of extant sample allele configurations can then be extended across time. This orthogonal polynomial approach is computationally efficient compared to the matrix multiplications of the discrete case. Again, changes in mutation-drift parameters can be accommodated, as well as demographic changes through concurrent rescaling of time. These modifications necessitate a change in the base of the orthogonal polynomials through linear transformation. Such transformations can cause numerical inaccuracies.

In Sec.~(\ref{section:particle}), we introduce our so-called particle model: Using coalescent arguments, one can trace the probability of sample configurations backward through the sample history similarly to the algorithm assuming complete linkage and the infinite sites model \citep{Fais15}. By conditioning on the extant sample, changing demography can be accounted for. Augmenting with temporal dynamics, the joint probability of the number of focal alleles in the sample and the past sample sizes can be determined for every point in time: Essentially, the history of the extant sample can be inferred. We show that the marginal distribution of the extant sample configuration derived using this backward particle approach is equivalent to that determined using the backward orthogonal polynomial approach. The sample genealogy can therefore be considered embedded in the spectral decomposition of the backwards diffusion generator similar to how it can be considered embedded within the discrete Moran model \citep[][chapt.~2.8]{Ethe11}. 

Assuming equilibrium and reversing time in the coalescent arguments, we obtain a forward particle approach that can also be used to determine the probabilities of sample configurations. It essentially corresponds to an urn model \citep{Step00} and is again similar to the algorithm assuming complete linkage equilibrium and the infinite sites model \citep{Wu10}. Recall that running the particle model forward in time means that past sample sizes become random variables and modelling non-equilibrium is not feasible. For a full forward-backward algorithm using the particle model, the backward particle approach must therefore be combined with the forward orthogonal polynomial approach. 

In Sec.~(\ref{section:particle_boundary}), we discuss different forward-backward algorithms for the boundary mutation-drift Moran model \citep{Vogl12}, which is a simplification of the general mutation Moran model for small scaled mutation rates. Inference, especially in non-equilibrium scenarios, becomes particularly efficient using orthogonal polynomials because the change of base required in the general mutation model can be avoided \citep{Vogl16}. The corresponding boundary mutation particle model, which we introduce in this article, allows for simple derivation of non-equilibrium transition probabilities. 

\section{Population allele proportions in the Moran and diffusion models}
\label{section:recap_Moran}

We assume a haploid population of size $N$ evolving according to a discrete space, bi-allelic, reversible, decoupled mutation-drift Moran model \citep[section 2.8]{Mora58a, Mora58b, Mora62,Ethe11}. Evolution proceeds stepwise, either by a birth-death event during which a randomly chosen individual is replaced the randomly chosen offspring of another or by an individual mutating. In Sec.~(\ref{section:discrete_Moran}) and Sec.~(\ref{section:diffusion_limit}) we establish our notation for the time discrete and diffusion versions of this Moran model respectively. The diffusion generator can be decomposed into eigenvectors (spectral decomposition) of modified Jacobi polynomials with corresponding eigenvalues. In other words, the transition density function has an explicit spectral representation; we review this in Sec.~(\ref{section:GeneralMutationDiffusion}). To infer population genetic parameters under these models, the marginal distributions of population allele frequencies have to be determined conditional on an extant sample. In the discrete case, a classic forward-backward algorithm on the transition rate matrix can be employed (Sec.~\ref{section:discrete_Moran_for_back}). In the diffusion limit, the Jacobi polynomials allow for efficient representation of the result (Sec.~\ref{section:diffusion_Moran_for_back}).

\subsection{The discrete decoupled Moran model}
\label{section:discrete_Moran}

\begin{figure}[!ht]
    \centering
    \includegraphics[width = 11.5cm]{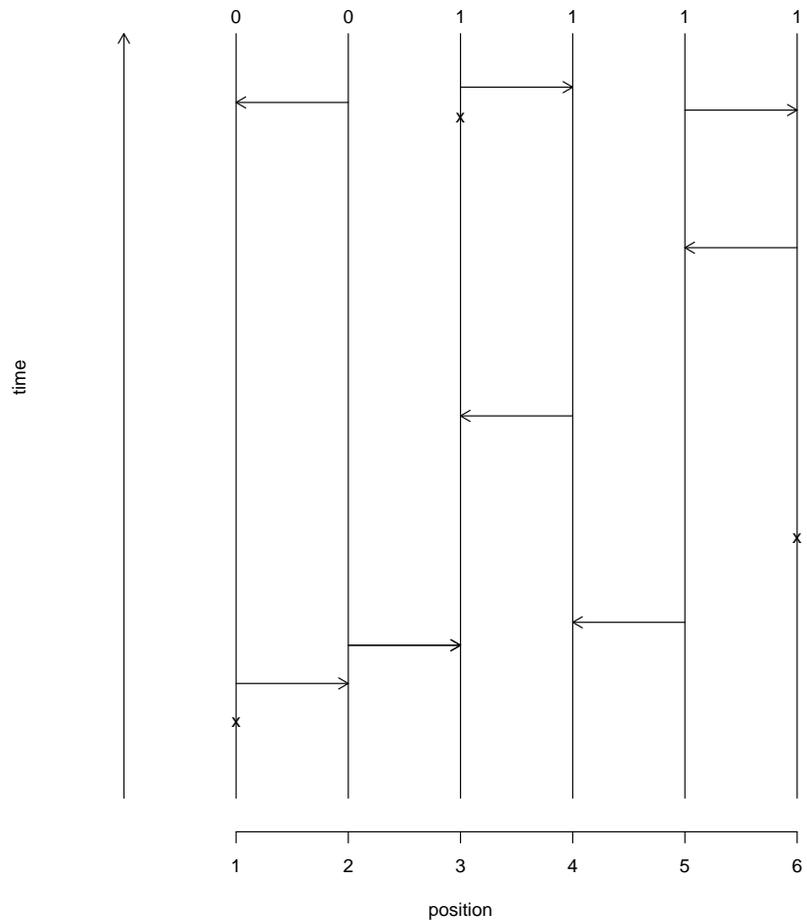}
    \caption{Schematic plot of a Moran model: The allelic type of the extant individuals is given at the top of the plot; backward in time the arrows indicate birth-death events and the 'x' mark mutation events. Note that the position and numbering of individuals on the x-axis is arbitrary.}
    \label{fig:Moran_general.eps}
\end{figure}

We consider a discrete time and discrete space, reversible, and decoupled mutation-drift Moran model with haploid population size $N$ \citep[][section 2.8]{Ethe11} to describe the evolution of the proportions of a focal allele $1$ and a non-focal allele $0$. We assume that birth-death events initially occur at rate $1$, while mutations arise at rates $\mu_1$ towards and $\mu_0$ away from the focal allele (so that the total mutation rate is $\mu=\mu_0+\mu_1$). We re-parameterize by setting the mutation bias towards the focal allele to $\alpha=\mu_1/(\mu_0+\mu_1)$ with $0<\alpha<1$, and equivalently define $\beta=1-\alpha$. Fig.~(\ref{fig:Moran_general.eps}) provides a visualisation for $N=6$ individuals, of which $y=4$ are of the focal type at the present time $s=0$.

\newpage
\newpage

Let $\x{s}$ ($0 \leq \x{s} \leq 1$) denote the relative frequency of allele $0$ at a non-focal locus at time $s$. The numbering of individuals is arbitrary. The transition rate matrix $\mathbf{T}_{i,j}$ is tridiagonal, aperiodic, and right stochastic:
\begin{equation}\label{eq:transition_decoupled_Moran}
\mathbf{T}_{i,j}=\Pr(\x{s+1}= \tfrac{j}{N} \given \x{s}= \tfrac{i}{N}) = \begin{cases}
\x{s}(1-\x{s})+\beta\mu \x{s} &\text{for j=i-1}\\
1-2 \x{s}(1-\x{s})-\beta\mu \x{s} - \alpha\mu \x{s} &\text{for j=i }\\
\x{s}(1-\x{s})+\alpha\mu \x{s} &\text{for j=i+1}\\
0 \qquad &\text{otherwise}\,.
\end{cases}
\end{equation}

Following convention, we re-scale the Moran events to bring the birth-death rate to $N^2$, and equivalently time becomes $t=s/N$ (note that this differs from the otherwise identical treatment in \citep[][formula~2.10]{Ethe11}, where time is scaled by $\binom{N}{2}$ to match the rate of the coalescent in the Wright-Fisher model). We also set the overall scaled mutation rate to $\theta=N(\mu_0+\mu_1)$. Then, the law of total probability allows us to write the forward N-particle generator of the Moran process as \citep[][formula~75]{Berg17}:
\begin{equation}\label{eq:forw_partly_cont_mutation}
\begin{split}
{\cal L}_N \Pr(N \x{t}) &= 
\alpha \theta \big((N-i+1)\Pr(N\x{t}=i-1)- (N-i)\Pr(N\x{t}=i)\big)\\
\qquad+ \beta \theta &\big((i+1)\Pr(N\x{t}=i+1)- i\Pr(N\x{t}=i)\big)\\
\qquad+ &\big((i-1)(N-i+1)\Pr(N\x{t}=i-1)\\
&+(i+1)(N-i-1)\Pr(N\x{t}=i+1) - 2i(N-i)\Pr(N\x{t}=i) \big)\,.
\end{split}
\end{equation}
Note that that the terms have been collected so that the first two summands account for mutation events and the remaining for genetic drift.

Suppose a sample of size $K$, with $0 \leq  K \leq N$, is drawn from the population at the current time $t=0$. The likelihood of observing $\y$ of the focal alleles in the sample follows a hypergeometric distribution:
\begin{equation}\label{eq:hypergeometric}
    \Pr(y \given K,N, Nx =i,t=0)= \frac{\binom{i}{y}\binom{N-i}{K-y}}{\binom{N}{K}}\,.
\end{equation}

Calculation of sample allele configurations at previous times requires the backward N-particle generator (note the generator ${\cal L}_N^{'}$ operating on the sample allele frequency in Eq.~(\ref{eq:hypergeometric}) below is defined equivalently to the forward generator ${\cal L}_N$ in Eq.~(\ref{eq:forw_partly_cont_mutation}) acting on the transition density from Eq.~(\ref{eq:transition_decoupled_Moran})):
\begin{equation}\label{eq:backw_partly_cont_mutation}
\begin{split}
&{\cal L}_N^{'}\Pr(\y \given K, Nx=i,t) = \\
&\qquad \alpha \theta (N-i) \bigg(\Pr(\y\given K, Nx=i+1, t)-\Pr(y\given K, Nx=i, t)\bigg)\\
&\qquad+ \beta \theta i \bigg(\Pr(\y\given K, Nx=i-1,t)-\Pr(\y\given K, Nx=i, t)\bigg)\\
&\qquad+ i(N-i) \bigg(\Pr(\y\given K, Nx=i+1,t)+(\Pr(\y\given K, Nx=i-1, t)\\
&\qquad\quad-2\Pr(\y\given K, Nx=i, t)\bigg)\,.
\end{split}
\end{equation}

\subsection{The diffusion limit}
\label{section:diffusion_limit}

Passing to the diffusion limit $N \to \infty$ is comparatively straightforward within the decoupled Moran model \emph{vs} the classic Moran model (see \citep[][chapt.~4]{Ewen04} for the derivation assuming the classic Wright-Fisher model, rescaling suffices to obtain equivalent results for the Moran model):
Set $\delta x=1/N$ and denote the transition rate density of continuous allele proportions as $\phi(x\given t)$. Then, the continuous forward N-particle generator is (compare: Eq.~(\ref{eq:forw_partly_cont_mutation}) \citep[][formula~76]{Berg17}:
\begin{equation}\label{eq:forw_partly_cont_mutation_2}
\begin{split}
{\cal L}_N\phi(x\given t) &=
\alpha\theta \bigg(\frac{(1-x+\delta x)\phi(x-\delta x\given t) - (1-x)\phi(x\given t)}{\delta x}\bigg)\\
&\qquad+\beta\theta \bigg(\frac{(x+\delta x)\phi(x+\delta x\given t) - x\phi(x\given t)}{\delta x}\bigg)\\
&\qquad+\bigg(\frac{(x-\delta x)(1-x+\delta x)\phi(x-\delta x\given t)}{\delta x^2}\\
&\qquad+ \frac{(x+\delta x)(1-x-\delta x)\phi(x+\delta x\given t)}{\delta x^2}-\frac{2x(1-x)\phi(x\given t)}{\delta x^2}\bigg).\\
\end{split}
\end{equation}
Taking $\lim_{N\to \infty} {\cal L}_N\phi(x\given t)$ and recognizing that each term defines a first or second derivative with respect to $x$ immediately recovers the infinitesimal operator \citep[][formula 2.11]{Ethe11}
\begin{equation}\label{eq:forw_operator}
{\cal L}\phi(x\given t) = -\frac{\partial}{\partial x}\theta(\alpha-x)\phi(x\given t) +\frac{\partial^2}{\partial x^2}x(1-x)\phi(x\given t)\,
\end{equation}
of the Kolmogorov forward (Fokker-Planck) diffusion equation for a general biallelic mutation-drift model:
\begin{equation}\label{eq:forw_mutdrift}
    \frac{\partial}{\partial t}\phi(x\given t)={\cal L} \phi(x\given t).
\end{equation}

Again, let us take a population sample of size $K$, $0 \leq K < \infty$, at the current time $t=0$. The likelihood of a sample of size $K$ with $\y$ alleles of the focal type is binomial:
\begin{equation}\label{eq:binomial_likelihood}
    \Pr(y\given K,x=\tfrac{i}{N},t=0)=\binom{K}{y}x^{y} (1-x)^{K-y}\,.
\end{equation}

The trajectory of the sample allele proportions backward in time is described by the Kolmogorov backward equation: 
\begin{equation}\label{eq:backw_mutdrift}
-\frac{\partial}{\partial t} \Pr(\y\given K, x,t)={\cal L}^{'}\Pr(\y\given K, x,t),
\end{equation}
with the backwards operator derived from Eq.~(\ref{eq:backw_partly_cont_mutation}) \citep[][formula~80]{Berg17}: \begin{equation}\label{eq:backw_operator}
 {\cal L}^{'}\Pr(\y\given K, x, t)=
    \theta(\alpha-x)\frac{\partial}{\partial x}\Pr(\y\given K, x, t)+x(1-x)\frac{\partial^2}{\partial x^2}\Pr(\y\given K, x, t)\,.
\end{equation}
The discrete probability distribution $\Pr(y\given K,x,t)$ in Eq.~(\ref{eq:backw_mutdrift}) is interpreted as the probability of obtaining the extant sample configuration $(y,K)$ conditional on an allele proportion $x$ at past times $t$ \citep{Berg17}. Note that the negative sign on the left side of the backward diffusion equation Eq.~(\ref{eq:backw_mutdrift}) (opposite to \citep{Ewen04}), ensures compatibility of the direction of time between the forward and backward Kolmogorov equations \citep{Zhao13a}.

\subsubsection{Modified Jacobi polynomials}
\label{section:GeneralMutationDiffusion}

Obtaining explicit analytical representations of the transition density at different time steps is a non-trivial problem. We will use the method of \citet{Song12} as adapted by \citet{Berg17} and introduce the (modified) Jacobi polynomials (compare also formula~22.3.2 in \citep{Abra70}):
\begin{equation}\label{eq:Jacobi_modified}
  R_n^{(\alpha,\theta)}(x)=\sum_{l=0}^n(-1)^l\frac{\Gamma(n-1+l+\theta)\Gamma(n+\alpha\theta)}{\Gamma(n-1+\theta)\Gamma(l+\alpha\theta)l!(n-l)!}x^l\,,
\end{equation}
where $n$, with $0\leq n\leq\infty$, is the order of the polynomial. Note that any  regular polynomial of order $n$ can be represented as a weighted sum of the above modified Jacobi polynomials. 

The modified Jacobi polynomials for any orders $m,n$, fulfill the following orthogonality relationship with respect to the weight function $w(x,\alpha,\theta)=x^{\alpha\theta-1}(1-x)^{\beta\theta-1}$: 
\begin{equation}\label{eq:ortho_Jacobi}
    \int_0^1 R_n^{(\alpha,\theta)}(x) R_m^{(\alpha,\theta)}(x)\, w(x,\alpha,\theta)\,dx=\delta_{n,m} \Delta_n^{(\alpha,\theta)}\,,
\end{equation}
where $\delta_{n,m}$ is Kronecker's delta, and
\begin{equation}
    \Delta_n^{(\alpha,\theta)}=\frac{\Gamma(n+\alpha\theta)\Gamma(n+\beta\theta)}{(2n+\theta-1)\Gamma(n+\theta-1)\Gamma(n+1)}\,
\end{equation}
is the proportionality constant.

The forward and backward Kolmogorov operators (Eqs.~(\ref{eq:forw_mutdrift}) and (\ref{eq:backw_mutdrift}), respectively) can be conveniently decomposed with the modified Jacobi polynomials as eigenfunctions (for details see \citep{Song12,Berg17}). The forward operator becomes:
\begin{equation}
    -\lambda_n w(x,\alpha,\theta) R_n^{(\alpha,\theta)}(x)={\cal L} w(x,\alpha,\theta) R_n^{(\alpha,\theta)}(x)\,,
\end{equation}
and the backward operator: 
\begin{equation}
    \lambda_n R_n^{(\alpha,\theta)}(x)={\cal L}^{'}R_n^{(\alpha,\theta)}(x)\,,
\end{equation}
with corresponding eigenvalues 
\begin{equation}\label{eq:eigenvalues}
    \lambda_n=n(n+\theta-1)\,.
\end{equation}

\subsection{Forward-backward algorithm}

Forward-backward algorithms are dynamic programming techniques that enable the efficient calculation of model states at any time from a sequence of observations. In our case, we have information from a population sample at the current time and aim to infer the distribution of past population allele frequencies. As shown in \citet{Berg17}, the forward-backward algorithm classically used for hidden Markov models \citep{RabinerJuang86, Vogl10} can be readily applied to the discrete decoupled Moran model of Sec.~(\ref{section:discrete_Moran}): the population allele proportions are considered `hidden' states and the sample allele configurations `emitted'. In Sec.~(\ref{section:diffusion_Moran_for_back}), we again follow \citet{Berg17} in establishing that a forward-backward algorithm can be constructed for the diffusion model of Sec.~(\ref{section:diffusion_limit}) if the diffusion model is represented using the modified Jacobi polynomials.

\subsubsection{Discrete Moran model}
\label{section:discrete_Moran_for_back}

We here reproduce the outline of the forward-backward algorithm for the discrete decoupled Moran model \citep{Berg17}.

\paragraph{Forward in time}
We start at time $s=S$ and assume that the allele proportions of the ancestral population are distributed according to an arbitrary distribution $\bs{\rho}(x)$. Recall that the beta distribution describes the allele proportions of a bi-allelic, general mutation Moran model in equilibrium \citep{Wrig31}. Multiplying with the binomial sampling likelihood and integrating over allele proportions results a beta-binomial compound distribution as marginal likelihood. For the biallelic Moran model, the following beta-binomial distribution with arbitrary mutation-drift parameters is therefore the standard prior $\bs{\rho}(x)$: 
\begin{equation}\label{eq:beta-binomial_prior}
\begin{split}
    \Pr(N x = i \given N,\alpha,\theta)\\
    &=\binom{N}{i}\,\frac{\Gamma(\theta)}{\Gamma(\alpha\theta)\Gamma(\beta\theta)}\frac{\Gamma(i+\alpha\theta)\Gamma(N-i+\beta\theta)}{\Gamma(N+\theta)}\,.
\end{split}
\end{equation}

Any starting distribution can be represented as a row vector of probabilities $\fv{S}= \bs{\rho}(x)$, where each entry corresponds to the probability of allele proportions being $0$, $1$,...,$N$. The probabilities of allele proportions at any time between $s=S$ and $s=0$ given our prior distribution, $\fv{s}=\Pr(N x_{s} \given \bs{\rho})$, can be determined via:
\begin{equation}
\fv{s+1} = \fv{s}\mathbf{T}_{i,j} \quad (S \le s < 0),
\end{equation}
where ${T}_{i,j}$ is the transition matrix defined in Eq.~(\ref{eq:transition_decoupled_Moran}).

At $s=0$, the entries of row vector $\bv{0_i}$ are given by the hypergeometric sampling scheme from Eq.~(\ref{eq:hypergeometric}) for each possible extant focal allele proportion between $0$ and $N$. The marginal likelihood of the observed sample allele frequency is then:
\begin{equation}\label{eq:marg_lh}
\begin{split}
\Pr(\y \given K,x,\bs{\rho})&=\fv{0}\bv{0}'\\
    &= \fv{S} \mathbf{T}_{i,j}^{|S|} \bv{0}'\,.
\end{split}
\end{equation}

\paragraph{Backward in time}

The same marginal likelihood may be obtained by recursing backwards from our sampling step at $s=0$ with initial probabilities $\bv{0_i}$ to the ancestral population state at $s=S$. Define entries of the row vector $\bv{s,i}=\Pr(y \given K ,N, Nx_s=i)$---they can be interpreted as the probability of the data given the population allele proportion at time $s$. We can recurse back in time by:
\begin{equation}
\begin{split}
\bv{s}' = \mathbf{T}_{i,j} \bv{s+1}' \quad (0 \geq s > S)\,,
\end{split}
\end{equation}
At $s=S$, we again obtain the marginal likelihood Eq.~(\ref{eq:marg_lh}).

\paragraph{Joint and conditional probabilities} At any time $s$, the joint probability of the population allele proportion $\x{s} = \tfrac{i}{N}$, and the sample allele frequency $y$ conditional on the starting distribution $\bs{\rho}$ is:
\begin{equation}\label{eq:joint_xy_discr}
\Pr(\x{s}=\tfrac{i}{N},y \given \bs{\rho}) = (\fv{s})_i (\bv{s})_i\,.
\end{equation}
Furthermore, the probability of the population allele proportions $\x{s} = i/N$ conditional on both the sample allele frequency and the starting distribution is:
\begin{equation}\label{eq:cond_x|y_discr}
\Pr(\x{s}=\tfrac{i}{N} \given y ,\bs{\rho}) = \frac{(\fv{s})_i (\bv{s})_i}{\fv{s}\bv{s}'}\,.
\end{equation}

\paragraph{Summary} The forward-backward algorithm with the Moran model conforms to the canonical situation \citep{RabinerJuang86}: At each time point, the population allele proportions comprise $N$ hidden states and the transition matrix $\mathbf{T}_{i,j}$ is of dimension $N\times N$. Conditional on an observed sample at the current time $s=0$ and a prior distribution on the ancestral allele configuration, the distribution of past and current population allele proportions can be determined. In population genetics, population demographic events are usually modeled to occur at a specific time in the past. Note that driving changes in mutation parameters can be incorporated into this approach by assuming different parameters for the prior distribution at $s=S$ than for the transition rate matrix and therefore the times $S < s\leq 0$. Further changes in population demography may be modeled by time-dependent transition matrices.

\subsubsection{Diffusion Model}
\label{section:diffusion_Moran_for_back}

We now adapt the forward-backward algorithm to the diffusion model described in Sec.~(\ref{section:diffusion_limit}) using the modified Jacobi polynomials from Sec.~(\ref{section:GeneralMutationDiffusion}).

\paragraph{Forward in time}
Suppose that at time $t=S$, with $S\leq 0$, the distribution of ancestral allele proportions is given by the arbitrary distribution $\bs{\rho}(x)=\phi(x\given t=S)$. This distribution is often assumed to be the beta equilibrium distribution for the bi-allelic Moran model \citep{Wrig31}:
\begin{equation}\label{eq:bb}
\bs{\rho}(x)=\frac{\Gamma(\theta)}{\Gamma(\alpha\theta)\Gamma(\beta\theta)}\,x^{\alpha\theta-1}(1-x)^{\beta\theta-1}\,.
\end{equation}
(Note that the weight function of the modified Jacobi polynomials is proportional to this beta distribution.) The allele proportions further forward in time, $\phi(x\given t,\bs{\rho})$, are determined by the forward diffusion equation Eq.~(\ref{eq:forw_mutdrift}). The solution to the forward equation can be represented using modified Jacobi polynomials; the ancestral allele proportion distribution is first expanded to:
\begin{equation}
    \bs{\rho}(x)=\sum_{n=0}^\infty \rho_n^{(\alpha,\theta)} R_n^{(\alpha,\theta)}(x),
\end{equation}
where the $\rho_n^{(\alpha,\theta)}$ are a possibly infinite number of Jacobi coefficients that depend on $\bs{\rho}(x)$. More explicitly:
\begin{equation}\label{eq:rho_coeffs}
\rho_n^{(\alpha,\theta)}= \frac{1}{\Delta_n^{(\alpha,\theta)}} \int_{0}^1 w(x,\alpha,\theta) R_n^{(\alpha,\theta)}(x)\bs{\rho}(x)\,dx\,.
\end{equation}
We then incorporate temporal dynamics and obtain the full solution: 
\begin{equation}\label{eq:forward_given_rho}
\phi(x\given t,\bs{\rho})=w(x, \alpha, \theta) \sum_{n=0}^\infty \rho_n^{(\alpha,\theta)} R_n^{(\alpha,\theta)}(x) e^{\lambda_n (S-t)}\,.
\end{equation}

\paragraph{Backward in time}

Backward in time, we again start with our sample: The binomial likelihood of the sampled allele proportions at time $t=0$ in Eq.~(\ref{eq:binomial_likelihood}) is expressed as a regular polynomial up to order $K$ with coefficients $$a_{j=y+i}(K,y)=(-1)^i\binom{K}{y}\binom{K-y}{i}$$ for $0\leq i\leq K-y$, and zero otherwise. 
Let $\mathbf{a}(K,y)$ be the vector of coefficients $a_{j}(K,y)$ and $\mathbf{R}^{(\alpha,\theta)}$ be the matrix of coefficients $R_n^{(\alpha,\theta)}(x)$. Note that this matrix is lower triangular. Then the binomial distribution can be uniquely expanded into Jacobi polynomials via the following linear algebraic equation:
\begin{equation}\label{eq:matr}
    \mathbf{d}^{(\alpha,\theta)}(K,y)=\mathbf{a}(K,y)\mathbf{R}^{(\alpha,\theta)}
\end{equation}
 Note that the triangular structure of $\mathbf{R}^{(\alpha,\theta)}$ obviates matrix inversion. Now the binomial sampling distribution from Eq.~\ref{eq:binomial_likelihood} can be rewritten:
\begin{equation}\label{eq:binom2jacobi}
   \Pr(y\given K,x,\alpha,\theta,t=0)=\binom{K}{y}x^y(1-x)^{K-y}=\sum_{n=0}^K d_n^{(\alpha,\theta)}(K,y) R_n^{(\alpha,\theta)}(x)\,.
\end{equation}
Further back, at times $t$ ($S \le t \le 0$), the distribution of sample proportions is given by:
\begin{equation}\label{eq:binom_cond_t}
    \Pr(y\given K,x,\alpha,\theta,t)=\sum_{n=0}^K d_n^{(\alpha,\theta)}(K,y) R_n^{(\alpha,\theta)}(x)e^{\lambda_n t}\,.
\end{equation}

Using the orthogonality of the Jacobi polynomials (Eq.~\ref{eq:ortho_Jacobi}), the continuous marginal likelihood becomes: 
\begin{equation}\label{eq:marg_like_general}
\begin{split}
\Pr(\y\given K,\alpha,\theta,S,\bs{\rho})&=  \int_0^1\phi(x\given t,\bs{\rho}) \Pr(y\given K,x, \alpha,\theta,t=0)\,dx\\
    &=\sum_{n=0}^K \rho_n^{(\alpha,\theta)} d_n^{(\alpha,\theta)}(K,y) \Delta_n^{(\alpha,\theta)} e^{\lambda_n S}\,. 
\end{split}
\end{equation}
Note that the expansion $\rho_n^{(\alpha,\theta)}$ may be infinite; however, calculation of the marginal likelihood only requires expansion to the order of the sample size $K$.

As briefly noted in the summary of Sec.~(\ref{section:discrete_Moran_for_back}), it is often convenient to be able to account for population demographics. We will consider this possibility for the diffusion approach: Assume the mutation parameters change from $\alpha$ to $\alpha^{*}$ and from $\theta$ to $\theta^{*}$ at time $t=S$. The coefficients of the new polynomial expansions can be obtained by linear transformation. More explicitly, consider a simple model where:
\begin{itemize}
\item{i)} at the present time $t=0$, the sample allele configuration $(y,K)$ is given; 
\item{ii)} between the times $S\leq t\leq 0$, the population genetic parameters $\alpha$ and $\theta$ remain constant; and 
\item{iii)} at time $t=S$ in the past the population allele proportion is beta distributed according to: 
\begin{equation} \label{eq:bb_time_change}
\bs{\rho}(x)=\frac{\Gamma(\theta^{*})}{\Gamma(\alpha^{*}\theta^{*})\Gamma(\beta^{*}\theta^{*})}\,x^{\alpha^{*}\theta^{*}-1}(1-x)^{\beta^{*}\theta^{*}-1}
\end{equation}
In fact, the allele proportion is beta distributed as above between $-\infty\leq t\leq S$ and can be expressed as the series $\rho_n^{(\alpha^{*},\theta^{*})}R_n^{(\alpha^{*},\theta^{*})}(x)$. 
\end{itemize}
The continuous marginal distribution then becomes: 
\begin{equation}\label{eq:marg_like_general_tc}
\begin{split}
\Pr(\y\given K,\alpha,\theta,S,\bs{\rho})&=  \int_0^1\phi(x\given t,\bs{\rho}) \Pr(y\given K,x,\alpha,\theta,t=0)\,dx\\
    &=\sum_{n=0}^K \rho_n^{(\alpha^{*},\theta^{*})} d_n^{(\alpha,\theta)}(K,y) \Delta_n^{(\alpha,\theta)} e^{\lambda_n S}\,. 
\end{split}
\end{equation}

\paragraph{Joint and conditional probabilities}
At any time $t$, the joint probability of the population allele proportion $\x{t} = \tfrac{i}{N}$, and the number of focal alleles in the sample $y$ conditional on the starting distribution $\bs{\rho}$ can be determined:
$$\Pr(\x{t} = \tfrac{i}{N}, y \given \bs{\rho})=\phi(x\given t,\bs{\rho}) \Pr(y\given K,x,t=0).$$
The probability of the population allele frequencies $\x{t} = \tfrac{i}{N}$ conditional on the both the sample allele proportions and the starting distribution is:
$$
\Pr(\x{t} = \tfrac{i}{N} \given y ,\bs{\rho})= \frac{\Pr(\x{t} = \tfrac{i}{N},  y \given \bs{\rho})}{\Pr(\y\given K,\alpha,\theta,S,\bs{\rho})}\,.
$$ 
We will not detail these equations here.

\paragraph{Summary} The forward-backward algorithm with the continuous diffusion mo\-del represented using modified orthogonal Jacobi polynomials deviates from the canonical situation: A transition kernel for population allele proportions is employed, which is expanded into an infinite-dimensional system of eigenfunctions and corresponding eigenvalues. For representing sample allele frequencies, however, only an expansion of the order of the sample size is needed. Indeed most problems only require a polynomial expansion up to the order of the sample size and the temporal system required is diagonal and thus extremely simple. Furthermore, a change in the mutation parameters can also be incorporated.

\section{Particle models and orthogonal polynomials}
\label{section:particle}
In this section, we introduce a novel forward-backward algorithm that harnesses together three components: 
\begin{itemize}
\item{i)} An approach based on orthogonal polynomials to describe the evolution of the 
population allele frequencies forward in time (Sec.~\ref{section:conditional_upsilon_kappa}),
\item{ii)} a so-called particle model that yields the conditional probabilities of the proportion of the focal allele at any point in the history of the sample by running traditionally backwards-looking coalescent arguments not only backward but also forward in time (Sec.~\ref{section:forward-particle}), (Sec.~\ref{section:backward-particle}); this is augmented by
\item{iii)} backward-in-time temporal dynamics accounting for the effect of changing mutation parameters on the sample sizes (Sec.~\ref{section:backward-particle_temp}) to yield probabilities of all past sample configurations. 

In total, we will arrive at the following joint probability:
\begin{equation}
    \Pr(\upsilon,\kappa\given K,y,\alpha,\theta,t)\,.
\end{equation}
This is the probability of seeing $\upsilon$ focal alleles in a sample of size $\kappa$ at any time $t<0$ in the past, conditional on an extant sample of size $K$ with $y$ focal alleles (and the underlying mutation-drift parameters $\theta$ and $\alpha$) at time $t=0$. 
\end{itemize}
We now begin by motivating the particle model.

\paragraph{Sample genealogy backward in time: coalescent}
In the (decoupled) Moran model of population size $N$, time is conventionally scaled so that the genealogies of all individuals in the population conform to the Kingman coalescent. Furthermore, the genealogy of any sample of size $K<N$ is embedded within the (decoupled) Moran model: The sample probabilities and transition rates of the coalescent remain unaffected by a change in sample size. In other words, the classic (decoupled) Moran model is sample consistent, and is dual to the Kingman coalescent in the sense that the expected population allele frequencies are identical between the two models \citep[][chapt.~2.8]{Ethe11}. 

\begin{figure}[!ht]
    \centering
    \includegraphics[width = 9cm]{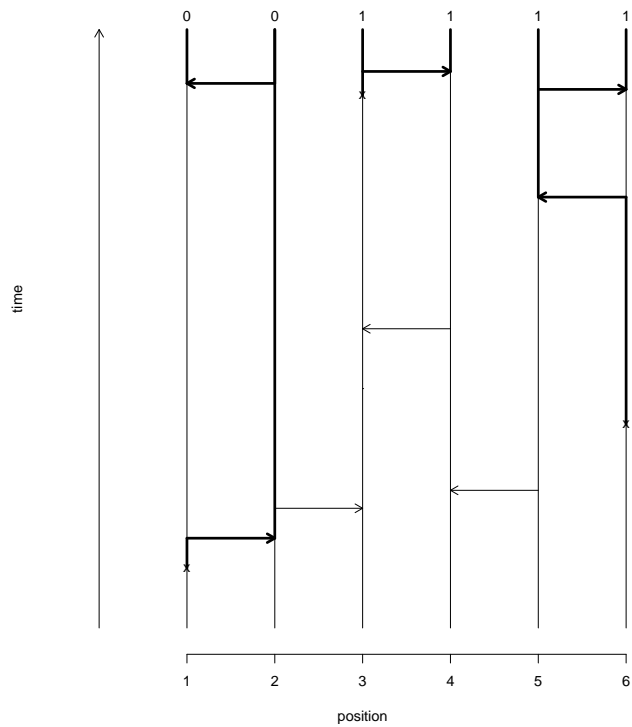}
    \caption{Schematic plot of a Moran model and embedded sample genealogy: The allelic type of the extant individuals is given at the top of the plot; backwards in time the arrows indicate birth/death events and the 'x' are mutation events. Note that the position and numbering of individuals on the lower x-axis is arbitrary. The bold lines and arrows correspond to the coalescent and mutation history of the sample respectively.}
    \label{fig:Moran_coal.eps}
\end{figure}

In Fig.~(\ref{fig:Moran_coal.eps}), the history of a sample of size $K=6$ with $y=4$ focal alleles is depicted. Both mutation and coalescent events are uninformative regarding the immediately preceding allelic state. Starting at the present time $t=0$ and looking back, the size of the sample $\kappa$ with $K\geq \kappa \geq 0$ is reduced in discrete stages by these events \citep[compare][for a similar argument in the context of the infinite sites model]{Fais15}. With the usual scaling of time, the rate of a coalescent is $\kappa(\kappa-1)$ and that of a mutation $\kappa\theta$, hence the total rate of reduction events is $\kappa(\kappa-1+\theta)$.
 At any time $t < 0$ the number of alleles $\kappa$ remaining from the original sample of size $K$ at $t=0$ is a random variable, as is the number of focal alleles remaining from the $y$ focal alleles at $t=0$, these we denote $\upsilon$. The probability of a reduction in the number of focal alleles from $\upsilon$ to $\upsilon-1$ in each backward time step is proportional to $\upsilon(\upsilon-1+\alpha\theta)$ (the contribution of the coalescent event is $\upsilon(\upsilon-1)$, that of the mutation event $\alpha\theta \upsilon$). The analogous reduction probabilities for the non-focal alleles are $(\kappa-\upsilon)(\kappa-\upsilon-1+\beta\theta)$. 
 Note that these reduction probabilities are for ordered events. However, we will treat allele 'labels' as interchangeable in each time step, and therefore use unordered reduction probabilities. We obtain these by dividing each of the previous ordered reduction probabilities by the number of potential alleles selected for a reduction event.
 
Let us now reverse the direction of time: Consider the indicator variable $z_{\kappa+1}$, which is one if the $\kappa+1$st allele is of the focal type and zero otherwise.
In each forward step, the probability of going from $\kappa$ to $\kappa+1$ unordered focal alleles is:
\begin{equation}\label{eq:ratio_general}
\begin{split}
 {\Pr}(z_{\kappa+1}\given \kappa,\upsilon,\alpha,\theta)&=\frac{(\upsilon+\alpha\theta)^{1-z_\kappa}(\kappa-\upsilon+\beta\theta)^{z_\kappa}}{\kappa+\theta}\\
&=\frac{\Gamma(\kappa+\theta)}{\Gamma(\kappa+1+\theta)}\frac{\Gamma(\upsilon+z_\kappa+\alpha\theta)}{\Gamma(\upsilon+\alpha\theta)}
\frac{\Gamma(\kappa+1-\upsilon-z_\kappa+\beta\theta)}{\Gamma(\kappa-\upsilon+\beta\theta)}\,.
\end{split}
\end{equation}

\newpage
\newpage

\paragraph{Sample genealogy forward in time: urn models}

Let us briefly view the probability of sample configurations running forward in time as an urn model similar to the Polya- or Hoppe-urns or the urn-models in \citep{Step00}. Hereto, we introduce a new time index $t_{\kappa-1}$ that runs from $t_0=0$, a time in the past at which only one individual from the sample is present in the population, to the present time $t_K$. The urn model is then given by the following algorithm: 
\begin{itemize}
    \item {Initiation.} Start with a sample of size $\kappa=1$ at time $t_0=0$. The probability of the focal allele is set to $\alpha$.   
    \item{Recursion.} Add a period of rate $\kappa(\kappa-1+\theta)$ to $t_{\kappa-1}$ obtain $t_{\kappa}$. Increase the sample size from $\kappa$ to $\kappa+1$ by a focal allele with probability
    \begin{equation}
        p_{\upsilon+1}=\frac{\upsilon + \alpha\theta}{\kappa+\theta}\,,
    \end{equation}
    or increase the sample size by a non-focal allele with probability $1-p_{\upsilon+1}$. Together this corresponds to the probability of the indicator variable in Eq.~(\ref{eq:ratio_general}). 
    \item{Stop.} When the sample size $\kappa=K$ is reached, add a period of rate  $K(K-1+\theta)$ to $t_{K-1}$ obtain $t_K$.
\end{itemize}
In order to obtain our usual time index with an extant time of zero, $t_K$ must be subtracted from all $t_{\kappa-1}$. Thus running the coalescent process forwards in time as in Eq.~(\ref{eq:ratio_general}) yields an urn model. 

\subsection{Particle model: algorithms}

We will now demonstrate how running coalescent arguments for the genealogy of an unordered sample from a decoupled Moran model both forward and backward in time can be used to determine sample allele frequencies for every past sample size (Sec.~\ref{section:forward-particle}), (Sec.~\ref{section:backward-particle}). Note that the past sample sizes are initially assumed given because they are determined by the constant rate of coalescence. Clearly, these sample sizes depend on mutation-drift parameters that are not necessarily constant across time. Note, however, that the number of focal alleles is conditionally independent of time given the (past) sample sizes $\kappa$ and the mutation-drift parameters. Hence we first treat the time independent dynamics, and then augment this with differential backward equations (Sec.~\ref{section:backward-particle_temp}) to account for changing mutation-drift parameters.

\subsection{Particle model: algorithm for forward probabilities}
\label{section:forward-particle}
For every $0\leq \kappa\leq K$, let $f(.,\kappa)$ denote a forward row vector of length $\kappa+1$. Each entry $f(\upsilon,\kappa)$ can be interpreted as a probability $\Pr(\upsilon\given \kappa,\alpha,\theta)$, where $0\leq \upsilon\leq \kappa$.
\begin{itemize}
    \item{Initiation.} Start with a sample of size $\kappa=0$. Trivially, the frequency of the focal allele is $\upsilon=0$, so set $f(\upsilon=0,\kappa=0)=1$.  
    \item{Recursion.} Move from sample size $\kappa$ to $\kappa+1$ by calculating every entry of $f(.,\kappa+1)$:
    \begin{equation}
    \begin{split}
     f(\upsilon,\kappa+1)&= \frac{\kappa-\upsilon-1+\beta\theta}{\kappa+\theta}\,f(\upsilon,\kappa-1)+\frac{\upsilon-1+\alpha\theta}{\kappa+\theta}\,f(\upsilon-1,\kappa-1)\,.
    \end{split}
    \end{equation}    
    \item{Stop.} Stop when the sample size $\kappa=K$ is reached. 
\end{itemize}
With time independent mutation-drift parameters, $f(\upsilon,\kappa)$ is a beta-binomial compound distribution:
\begin{equation}\label{eq:beta-binomial}
\begin{split}
    f(\upsilon,\kappa)&=\Pr(\upsilon\given\kappa,\alpha,\theta)\\
    &=\binom{\kappa}{\upsilon}\,\frac{\Gamma(\theta)}{\Gamma(\alpha\theta)\Gamma(\beta\theta)}\frac{\Gamma(\upsilon+\alpha\theta)\Gamma(\kappa-\upsilon+\beta\theta)}{\Gamma(\kappa+\theta)}\,.
\end{split}
\end{equation}

\subsection{Particle model: algorithm for backward probabilities}
\label{section:backward-particle}
For every $0\leq \kappa\leq K$, introduce the backward row vector $b(.,\kappa)$ of length $\kappa+1$. Every entry $b(\upsilon,\kappa)$ corresponds to ${\Pr}(y\given K,\kappa,\upsilon,\alpha,\theta)$ for $0\leq \upsilon\leq \kappa$, which is the probability of the observed number of focal alleles given past allele configurations. 
\begin{itemize}
    \item {Initiation.} Start with a sample of size $K$ with $y$ focal alleles. Set $\upsilon=y$ and $\kappa=K$, and therefore $b(\upsilon=y,\kappa=K)=1$.  
    \item{Recursion.} Move from $\kappa+1$ to $\kappa$ by calculating the entries of $b(.,\kappa)$:
    \begin{equation}
    \begin{split}
     b(\upsilon,\kappa)&= \frac{\kappa-\upsilon+\beta\theta}{\kappa+\theta}\,b(\upsilon,\kappa+1)+\frac{\upsilon+\alpha\theta}{\kappa+\theta}\,b(\upsilon+1,\kappa+1)\,.
    \end{split}
    \end{equation}    
    \item{Stop.} End the recursion, when the sample size $\kappa=0$ is reached. Note that $b(\upsilon=0,\kappa=0)={\Pr}(y\given K,\upsilon=0,\kappa=0,\alpha,\theta)$ corresponds to the likelihood ${\Pr}(y\given K,\alpha,\theta)$.
\end{itemize}
With time independent mutation-drift parameters, $b(\upsilon,\kappa)$ is again a beta-binomial compound distribution:
\begin{equation}\label{eq:beta_binom_upsilon_kappa}
\begin{split}
    b(\upsilon,\kappa)&=\Pr(y\given K,\kappa,\upsilon,\alpha,\theta)\\
    &=\binom{K-\kappa}{y-\upsilon}\,\frac{\Gamma(\kappa+\theta)}{\Gamma(\upsilon+\alpha\theta)\Gamma(\kappa-\upsilon+\beta\theta)}\frac{\Gamma(y+\alpha\theta)\Gamma(K-y+\beta\theta)}{\Gamma(K+\theta)}\,.
\end{split}
\end{equation}

\subsection{Particle model: time independent conditional probabilities of focal alleles}

\begin{figure}[!ht]
    \centering
    \includegraphics[width = 11.5cm]{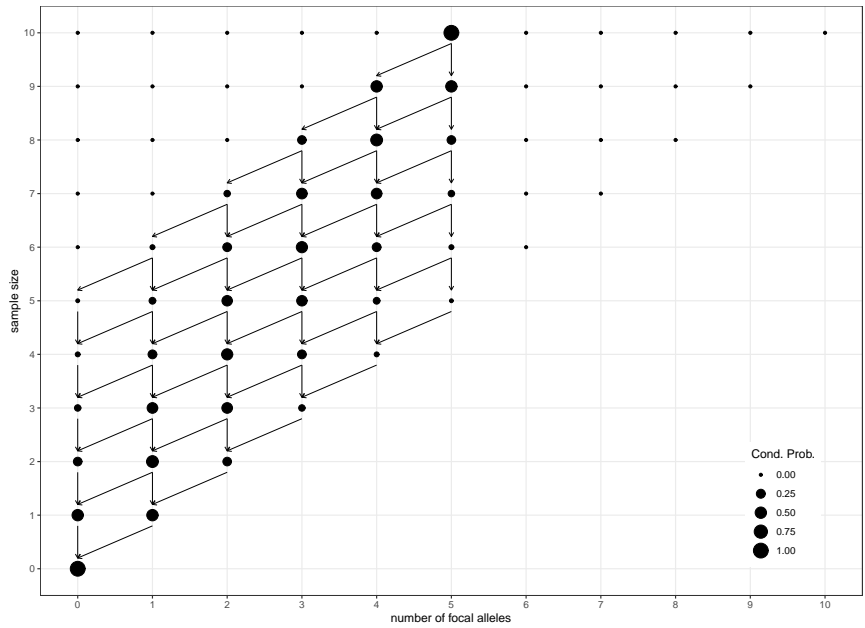}
    \caption{Consider a sample of size $K = 10$ with $y=5$ focal alleles, \ie\ a starting configuration of $(5,10)$. Assume a mutation bias of $\alpha = 0.3$ towards the focal allele and an overall scaled mutation rate of $\theta = 0.1$. Here, we show the conditional probabilities of all possible particle configurations between $(5,10)$ and $(0,0)$ as per Eq.~(\ref{eq:conditional_upsilon_given kappa}). This is an application of the forward- backward particle model in Secs.~(\ref{section:forward-particle}) and (\ref{section:backward-particle}).
}
    \label{fig:num_example_equilibrium}
\end{figure}

We can now combine the forward and backward algorithms and calculate the likelihood of seeing $y$ focal alleles in a sample of size $K$ given time independent mutation-drift parameters (note the use of Vandermonde's identity to obtain the final result):
\begin{equation}
\begin{split}
    \Pr(y\given K,\alpha,\theta)&=\sum_{\upsilon=0}^\y f(\upsilon,\kappa)b(\upsilon,\kappa)\\
    &=\sum_{\upsilon=0}^\y
    \binom{\kappa}{\upsilon}\,\frac{\Gamma(\theta)}{\Gamma(\alpha\theta)\Gamma(\beta\theta)}\frac{\Gamma(\upsilon+\alpha\theta)\Gamma(\kappa-\upsilon+\beta\theta)}{\Gamma(\kappa+\theta)}\\
    &\quad\times\binom{K-\kappa}{y-\upsilon}\,\frac{\Gamma(\kappa+\theta)}{\Gamma(\upsilon+\alpha\theta)\Gamma(\kappa-\upsilon+\beta\theta)}\frac{\Gamma(y+\alpha\theta)\Gamma(K-y+\beta\theta)}{\Gamma(K+\theta)}\\
    &=\sum_{\upsilon=0}^\y\binom{\kappa}{\upsilon}\binom{K-\kappa}{y-\upsilon} \frac{\Gamma(\theta)}{\Gamma(\alpha\theta)\Gamma(\beta\theta)}\frac{\Gamma(y+\alpha\theta)\Gamma(K-y+\beta\theta)}{\Gamma(K+\theta)}\\
    &=\binom{K}{y} \frac{\Gamma(\theta)}{\Gamma(\alpha\theta)\Gamma(\beta\theta)}\frac{\Gamma(y+\alpha\theta)\Gamma(K-y+\beta\theta)}{\Gamma(K+\theta)}\,.
\end{split}
\end{equation}
Similarly, the probability of seeing $\upsilon$ focal alleles in a given past sample of size $\kappa$ assuming a current sample of size $K$ containing $y$ focal alleles and time independent mutation-drift parameters is:
\begin{equation}\label{eq:conditional_upsilon_given kappa}
    \Pr(\upsilon\given K,y,\kappa,\alpha,\theta)=
    \frac{f(\upsilon,\kappa)b(\upsilon,\kappa)}{\sum_{\upsilon=0}^\kappa f(\upsilon,\kappa)b(\upsilon,\kappa)}\,.
\end{equation}
We provide a visual example of the likelihood of the focal locus traced backward in time in Fig.~(\ref{fig:num_example_equilibrium}).

\newpage

\newpage

\subsection{Particle model: temporal dynamics}
\label{section:backward-particle_temp}
So far, we have always assumed the past sample sizes $\kappa$ as given. However, they can easily be modelled as a pure death process dependent on mutation and drift: Both events simply reduce the sample size by one. 

The following system of differential equations describes the evolution of the sample size $\kappa$ backward in time: 
\begin{equation}\label{eq:temp_system_back}
\begin{split}
    -\frac{d}{dt}\Pr(\kappa=K\given  K,\theta,t)&=\lambda_K\,\Pr(\kappa=K\given  K,\theta,t)\,, \quad\text{and}\\ 
    -\frac{d}{dt}\Pr(\kappa\given  K,\theta,t)&=-\lambda_{\kappa+1}\,\Pr(\kappa+1\given  K,\theta,t)\\
    &\qquad+\lambda_{\kappa}\,\Pr(\kappa\given  K,\theta,t)\,, \quad\text{for $K>\kappa\geq 0$.}
\end{split}
\end{equation}
with corresponding eigenvalues  $\lambda_\kappa=\kappa(\kappa-1+\theta)$ (as in Eq.~(\ref{eq:eigenvalues})). 
The starting condition is: $\Pr(\kappa=K\given K,\theta, t=0)=1$, and  $\Pr(K>\kappa\geq 0\given  K,\theta,t=0)=0$. Then the solution of the above system of equations is \citep[][chapter 6, Eq.~2.2]{TaylorKarlin98}
\begin{equation}\label{eq:temp_system_back_solution}
\begin{split}
    \Pr(\kappa=K\given K,\theta,t)&=e^{\lambda_K t}\,,\\
    \Pr(\kappa\given K,\theta,t)&=\sum_{i=\kappa}^{K}c_{i,\kappa}\,e^{\lambda_i t}\, \text{, for $K-1\geq \kappa\geq 0$}\\
    &\text{with $c_{i,\kappa}=\frac{\prod_{j=\kappa+1}^K \lambda_j}{\prod_{j=\kappa,j\neq i}^K(\lambda_j-\lambda_i)}$.}
\end{split}
\end{equation}

\subsection{Particle model: total backward dynamics}

Augmenting the time independent backward variables from Eq.~(\ref{eq:beta_binom_upsilon_kappa}) with the temporal dynamics from the previous subsection, \ie\  Eq.~(\ref{eq:temp_system_back_solution}), we obtain the joint probability of the number of focal alleles in the extant sample and the past sample sizes:
\begin{equation}\label{eq:beta_binom_upsilon_kappa_t}
\Pr(y,\kappa\given K,\upsilon,\alpha,\theta,t)=
    \Pr(y\given K,\kappa,\upsilon,\alpha,\theta)\Pr(\kappa\given K,\theta,t)\,.
\end{equation}

\subsection{Particle model: joint and marginal probabilities}
\label{joint_y_upsilon_kappa}

The aim of the forward-backward algorithm is to obtain the probability of ancestral sample configurations $(\upsilon,\kappa)$ at arbitrary times $t$, conditional on the sample configuration $(y,K)$ at time $t=0$, \ie{} $\Pr(\upsilon,\kappa\given K,y,\dots)$. Multiplying the probability in Eq.~(\ref{eq:beta_binom_upsilon_kappa_t}), $\Pr(y,\kappa\given K,\upsilon,\dots)$, with the probability of the frequency of the focal allele $\upsilon$ given a past sample size $\kappa$ and population proportion $x$ gives us the joint probability $\Pr(y,\upsilon,\kappa\given K,x,\dots)$. Recall that the probabilities in Eq.~(\ref{eq:beta_binom_upsilon_kappa_t}) are for unordered samples, since we treat the allele 'labels' as interchangeable in each time step within the sample genealogy. Obtaining the number of focal alleles $\upsilon$ in an ancestral sample from a given population allele proportion $x$, however, requires ordered sampling (instead of binomial \ie{} unordered sampling):
\begin{equation}\label{eq:ordered_bin}
\Pr(\upsilon\given \kappa,x)^o=x^\upsilon(1-x)^{\kappa-\upsilon}\,,
\end{equation}
to obtain:
\begin{equation}\label{eq:joint_dist}
\Pr(y,\upsilon,\kappa\given K,x,\alpha,\theta,t)=
     \Pr(y\given K,\kappa,\upsilon,\alpha,\theta)\Pr(\upsilon\given \kappa,x)^o\Pr(\kappa\given K,\theta,t)\,.
\end{equation}

The marginal distribution of the number of focal alleles $y$ in the extant sample of size $K$ is then determined by:
\begin{equation}\label{eq:marg_particle}
    \Pr(y\given K,x,\alpha,\theta,t)= \sum_{\kappa=0}^{K}\sum_{\upsilon=\max(\kappa+y-K,0)}^{\min(\kappa,y)} \Pr(y,\upsilon,\kappa\given K,x,\alpha,\theta,t)\,,
\end{equation}
for every $t<0$. This is a sum over all possible $(K+1)(y+1)$ allele configurations. For $t\to -\infty$, the sum becomes dominated by the probability for the case $\kappa=0$ since the likelihood in Eq.~(\ref{eq:ordered_bin}) is trivially one for $\kappa=0$ irrespective of $x$. Hence, this can be used to obtain the likelihood:
\begin{equation}
\Pr(y\given K,x,\theta,\alpha,t\to\infty)=
     \Pr(y\given K,\theta,\alpha)\,.
\end{equation}

\subsubsection{Relationship between particle probabilities and Jacobi polynomials}
\label{section:particle_jacobi}

Recall that we previously obtained an expression for Eq.~(\ref{eq:marg_particle}) in terms of modified Jacobi polynomials (Eq.~\ref{eq:binom_cond_t}), which we reproduce here:
\begin{equation}\label{eq:orthopolynomial}
    \Pr(y\given K,x,\alpha,\theta,t)=\sum_{n=0}^K e^{\lambda_n t} d_n(K,y)^{(\alpha,\theta)} R_n^{(\alpha,\theta)}(x)\,.
\end{equation}
Clearly, these two representations must be equivalent: The order of expansion of the modified Jacobi polynomials is generally equivalent to the sample size, so we can set $\kappa$ to $n$ in Eq.~(\ref{eq:marg_particle}). Now, for each past sample size $\kappa$ (or now $n$), the terms with the same temporal component $e^{\lambda_n t}$ and the same power of $x$ from Eq.~(\ref{eq:marg_particle})(left below) and Eq.~(\ref{eq:orthopolynomial})(right below) can be equated:
\begin{equation}
\begin{split}
    e^{\lambda_n t}x^n c_{n,n}\sum_{\upsilon=\max(n+y-K,0)}^{\min(\kappa,y)}(-1)^{n-\upsilon}b(\upsilon,n)&= e^{\lambda_n t}x^n d_n(K,y)^{(\alpha,\theta)} r_{n,n} \\
     c_{n,n}\sum_{\upsilon=\max(n+y-K,0)}^{\min(\kappa,y)}(-1)^{n-\upsilon}b(\upsilon,n)&=  d_n(K,y)^{(\alpha,\theta)} r_{n,n} \,
\end{split}
\end{equation}
where the $R_n^{(\alpha,\theta)}(x)$ are written as the term $x^n$ multiplied by the corresponding coefficient $r_{n,l}$, and $c_{n,n}$ are the $n$th coefficients solving the system of temporal differential equations in Eq.~(\ref{eq:temp_system_back}). For $y=0$ or $y=K$, the above formula simplifies as there only a single term remains in the summation.

\subsection{Past allele configurations}
\label{section:conditional_upsilon_kappa}

In order to determine the distribution of allele configurations at any given point in time conditional on a current sample, we must combine the total backwards dynamics of a sample given in Eq.~(\ref{eq:beta_binom_upsilon_kappa_t}) with repeated sampling from the general population forward in time. Note that the latter cannot be modelled using a particle model: The sample sizes in the particle model - both current and past - are assumed fixed until augmented backward in time by temporal dynamics accounting for mutation-drift parameters; importantly, these temporal components are conditional on the extant sample size. When sampling from the population forward in time and agnostic to either past or future sample sizes and configurations, however, the sample size $\kappa$ becomes a random variable. 

The population allele proportion can be modelled forward in time by expanding the transition density $\phi(x\given t,\bs{\rho})$ into orthogonal polynomials as in Eq.~(\ref{eq:forward_given_rho}); $\bs{\rho}$ here represented the beta equilibrium distribution of the bi-allelic Moran model.
Further, the likelihood of observing $\upsilon$ alleles of the focal type in a sample of size $\kappa$ conditional on the population allele frequency $x$ can also be expanded into the orthogonal polynomials: 
\begin{equation}
    \Pr(\upsilon\given \kappa,x)=\binom{\kappa}{\upsilon} x^{\upsilon}(1-x)^{\kappa-\upsilon}=\sum_{n=0}^\kappa {d}_n(\kappa,\upsilon) R_n^{(\alpha,\theta)}(x)\,.
\end{equation}
Then the joint probability of the focal alleles $y$ in the extant sample and the past particle configurations becomes: 
\begin{equation}\label{eq:joint_y_i}
\begin{split}
    &\Pr(y,\upsilon,\kappa\given K,\alpha,\theta,t,\bs{\rho})\\
     &\qquad=\Pr(y,\kappa\given K,\upsilon,\alpha,\theta,t) \int_0^1 \Pr(\upsilon\given \kappa,x)\,\phi(x\given t,\bs{\rho})\,dx\\
    &\qquad=\Pr(\upsilon\given K,y,\kappa,\alpha,\theta)\,\Pr(\kappa\given K,\theta,t)\\
    &\qquad\qquad \times\sum_{n=0}^\kappa \,e^{\lambda_j(S-t)}\int_0^1 {d}_n(\kappa,\upsilon)  R_n^{(\alpha,\theta)}(x) \rho_j^{(\alpha,\theta)} R_n^{(\alpha,\theta)}(x)
     x^{\alpha\theta-1}(1-x)^{\beta\theta-1}\,dx\\
   &\qquad= \Pr(\upsilon\given K,y,\kappa,\alpha,\theta)\,\Pr(\kappa\given K,\theta,t)\,\sum_{n=0}^\kappa e^{\lambda_j(S-t)} {d}_n(\kappa,\upsilon)\rho_n^{(\alpha,\theta)} \Delta_n^{(\alpha,\theta)}\,.
\end{split}
\end{equation}
Note that the first term can be calculated using our forward-backward particle algorithm in Eq.~(\ref{eq:conditional_upsilon_given kappa}), the second is the solution of the temporal system in Eq.~(\ref{eq:temp_system_back_solution}).

Finally, the probability of ancestral allele configurations at any time, conditional on an extant sample, becomes:
\begin{equation}\label{eq:joint_y_i_2}
\begin{split}
    \Pr(\upsilon,\kappa \given K,y, \alpha,\theta,t,\bs{\rho})&=\frac{\Pr(y,\upsilon,\kappa\given K,\alpha,\theta,t,\bs{\rho})}{\Pr(y\given K,\alpha,\theta,t,\bs{\rho})}\,,
\end{split}
\end{equation}
where the denominator can be determined using either
\begin{itemize}
\item{(i)} the particle approach via 
$$\Pr(y\given K,\alpha,\theta,t,\bs{\rho})=\Pr(y\given K,x,\alpha,\theta,t)\int_0^1\phi(x\given t,\bs{\rho})\,dx\\$$
with the first term from Eq.~(\ref{eq:marg_particle}), or 
\item{(ii)} the equivalent polynomial approach in Eq.~(\ref{eq:marg_like_general}).
\end{itemize}

\begin{figure}[!ht]
    \centering
    \includegraphics[width = 9cm,angle=-90]{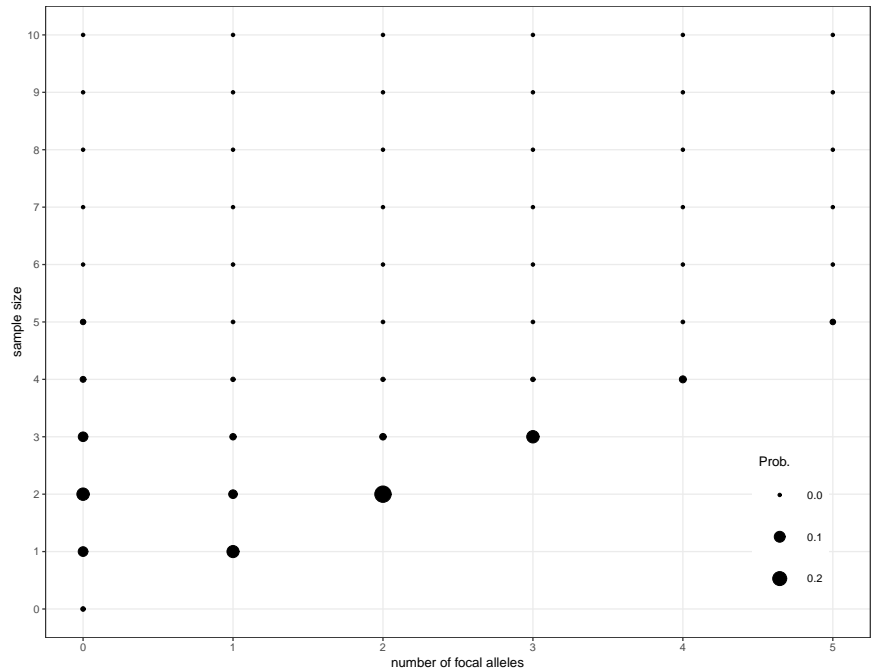}
    \caption{Let us again consider a sample of size $K=10$ with $y=5$ focal individuals drawn at the present time $t=0$. We now consider a non-equilibrium situation where the population size changes at $t=S=-0.5$ in the past; this change is driven by a single change in the scaled mutation rate: The ancestral rate $\theta^{*}=0.3$ changes to the current rate $\theta^{*}=0.1$, $\alpha=0.3$ remains constant throughout. We show the probabilities of the various possible allele configurations $\upsilon$ and $\kappa$ from Eq.~(\ref{eq:joint_y_i});
    the size of the dots is proportional to the probability of the configuration.}
    \label{fig:num_example_non-equilibrium}
\end{figure}

Changes in mutation parameters over time can also be incorporated in this forward-backward algorithm: Once again consider the scenario in which the mutation parameters $\alpha$ and $\theta$ remain constant between the times $S\leq t\leq 0$, and at time $t=S$ in the past they change to $\alpha^*$ and $\theta^*$. Recall that the ancestral allele configuration $\bs{\rho}$ is then modelled as beta-binomial with parameters $\alpha^*$ and $\theta^*$ instead of $\alpha$ and $\theta$ (Eq.~(\ref{eq:bb_time_change})), and this can be substituted into the equations of this subsection accordingly (directly in Eq.~(\ref{eq:joint_y_i}) and (i), and see Eq.~(\ref{eq:marg_like_general_tc}) for (ii)). An example is shown in Fig.~(\ref{fig:num_example_non-equilibrium}).

\paragraph{Summary} The forward-backward algorithm using a combination of the particle model and orthogonal polynomials deviates considerably from the canonical situation: Within the forward-backward recursions tracing the sample allele configurations described by the particle model, each step leads to a change in the sample size $\kappa$ and hence the dimension of the transition matrix, which is no longer square. Furthermore, the backwards temporal system accounting for the effect of changing mutation-drift parameters on the sample size is not diagonal and thus moderately complex. Calculating the marginal likelihood of the number of focal alleles in the extant sample involves a summation over $(K+1)(y+1)$ allele configurations rather than an eigensystem decomposition as with the orthogonal polynomials---this may sometimes be convenient though it is generally less efficient. However, the two methods yield equivalent results, which is of considerable theoretical interest. To determine the likelihood of any past population allele frequencies, the particle model is combined with forward population dynamics represented as orthogonal polynomials.   

\newpage
\newpage

\section{Transition probabilities of the boundary mutation-drift Moran model}
\label{section:particle_boundary}

Recall that in synonymous sites of protein coding genes of eukaryotes, the expected heterozygosity, which is roughly equal to the scaled mutation rate, is $\leq 10^{-2}$ \citep{Lynch06}. Note that the expected heterozygosity corresponds to the expected polymorphism in a sample of size two and is thus proportional to the number of polymorphic sites in the sample; increasing the sample size leads to a roughly logarithmic increase in the proportion of polymorphic sites. Population genetic models derived in the limit of small scaled mutation rates, $\theta\to 0$, often approximate general mutation dynamics with sufficient accuracy and have clear numeric advantages \citep{Vogl20}. A first order Taylor expansion in $\theta$ of the beta-binomial equilibrium distribution of a sample from the decoupled bi-allelic Moran model yields the equilibrium distribution of the proportion of alleles $\mathbf{X}$ at any locus in the so-called boundary mutation-drift Moran model \citep{Vogl12}:
\begin{equation}\label{eq:eq_boundary}
 {\Pr}(\mathbf{X}=\frac{i}{N}\given N,\alpha,\theta)=\begin{cases}
 \beta\big(1-\alpha\theta H_{N-1}\big) &\text{for $i=0$,}\\
 \alpha\beta\theta\,\frac{N}{i(N-i)} &\text{for $1\leq i\leq (N-1)$,}\\
 \alpha\big(1-\beta\theta H_{N-1}\big) &\text{for $i=N$}\,,
 \end{cases}
\end{equation}
where $H_{N-1}=\sum_{i=1}^{N-1}\frac{i}{N}$ is the harmonic number.
In essence, the boundary mutation-drift Moran model assumes that only a single mutation segregates in a population at any given time: Mutations arise exclusively from the monomorphic boundaries rather than from near the boundaries as in the general mutation model with low mutation rates; the polymorphic interior is governed by drift (and selection in non-neutral scenarios \citep{VoglMikula21}). Simulations show that the boundary mutation-drift Moran model is a good approximation for the general mutation Moran model if the expected equilibrium heterozygosity $2\alpha\beta\theta \leq 10^{-2}$ \citep{Vogl12}. Importantly, the boundary mutation-drift Moran model  simplifies inference considerably \citep{Vogl14}.

Several approaches have been taken to derive transition probabilities compatible with this equilibrium distribution \citep{Vogl16,BurdenGriffiths19,VoglMikula21}. \citet{BurdenGriffiths19} start their derivations from a Wright-Fisher diffusion with $\theta\to 0$, and obtain the probability of the admissible sample configurations under the constrained mutation rate at any time $t$ by considering all coalescent sub-trees and their scaled branch lengths. The resulting formulae are complex and not easily employed in an inferential framework. 

\begin{figure}[!ht]
    \centering
    \includegraphics[width = 10cm]{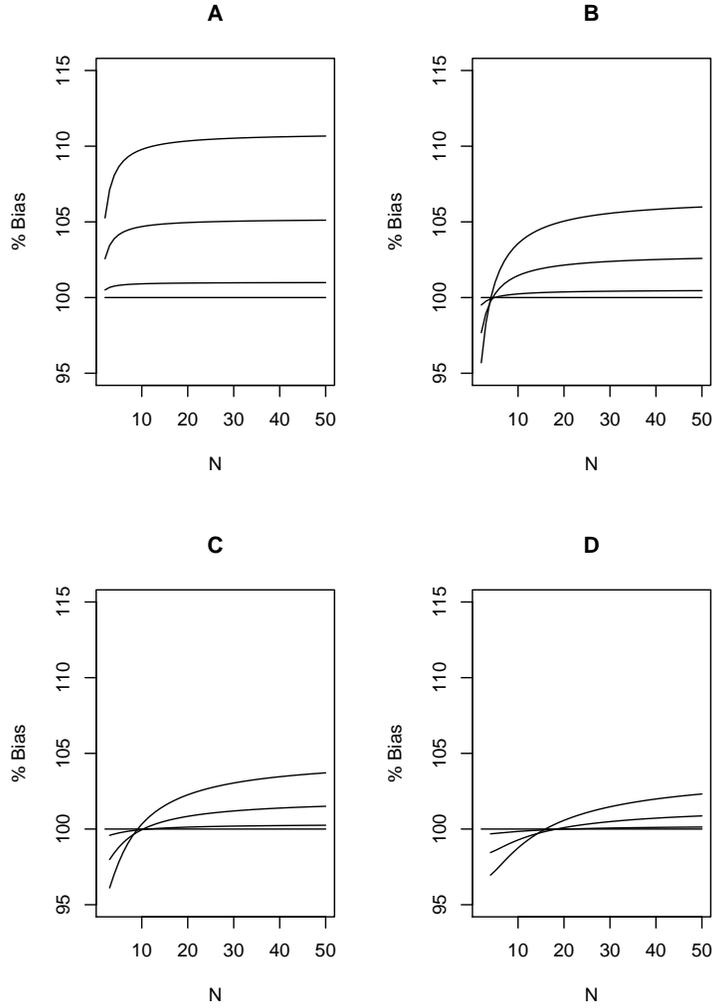}
    \caption{Percentage bias of the eigenvalues of the boundary mutation-drift Moran model vs. the eigenvalues of the general mutation Moran model in dependence on the haploid population size $N$: Each plot shows a different eigenvalue - A) $\lambda_1$, B) $\lambda_2$, C) $\lambda_3$, D) $\lambda_4$ - of the transition matrix with three different overall scaled mutation rates:  $\theta=c(0.1,0.05,0.01)$, whereby the largest overall scaled mutation rate corresponds to the slope showing the highest percentage bias, etc. In all plots, the reference line at 100\% corresponds to the value of the eigenvalue in the limit $\theta\to 0$.}
    \label{fig:eigenvalue}
\end{figure}

The transition rate matrix of the boundary-mutation Moran model given by \citet{VoglMikula21} is comparatively tractable. To balance the constraint that mutations are only allowed from the boundaries, mutation rates are re-scaled to obtain an equilibrium distribution corresponding to the Taylor series expansion of the general mutation Moran model. Importantly, this transition matrix and in particular its eigenvalues are not consistent with varying $N$, precisely because it is scaled so that mutations segregate at an identical average rate in equilibrium regardless of the sample size. Therefore the embedded genealogies of alleles depend on the sample size. For small scaled mutation rates the deviation of the eigenvalues from those of the general mutation Moran model (which are consistent for varying $N$) is negligible, as shown in Fig.~(\ref{fig:eigenvalue}). For larger values such as $\theta=0.025$, which is close to the limit of feasibility for the first order $\theta$ approximation, the first non-zero eigenvalue $\lambda_1$, which determines the asymptotic speed of approach to equilibrium, is increased by approximately $2.5\%$ to $5\%$. This could become important in a phylogenetic context, as split times between species may considerably exceed the (effective) population size $N$, which in turn constrains the effect of drift. 

The stationary distribution of the phylogenetic rendition of the boundary mutation-drift Moran model has traditionally been scaled to maintain the proportion of monomorphic sites across different sample sizes \citep{Schrempf16} (but note that one version can be reparametrized to give the other), and the corresponding transition matrix \citep{Borges19} is therefore similarly inconsistent. (Note, however, that with a given population size $N$ this approach provides a statistically consistent framework for inference \citep{Borges20}.) Divergence estimators based on branch lengths \citep{Schrempf16, Borges19} may be affected by this; note that this also depends on whether the deviation actually exceeds the margin of numerical accuracy of the implementation, which may not be the case. Determining substitution rates by multiplying the mutation rates and fixation/hitting probabilities derived from the continuous approximations avoids the problem entirely \citep{VoglMikula21}. 

\newpage

\newpage

An alternative approach to representing the transition rates of the boundary mutation-drift Moran model has been proposed by \citet{Vogl16}: In the limit of the small scaled mutation rate $\theta\to0$, the interior of the forward (Fokker-Planck) diffusion equation in Eq.~(\ref{eq:forw_mutdrift}) becomes a pure drift diffusion model that can be decomposed into eigenvectors given by (modified) Gegenbauer polynomials and corresponding eigenvalues $\lambda_n=n(n-1)$ for $n\geq 2$, where $0 \leq n \leq \infty$ is the order of the polynomials. The (modified) Gegenbauer polynomials can be obtained from the (modified) Jacobi polynomials by a Taylor series expansion up to zeroth order \citep{Vogl16}. To incorporate boundary dynamics similar to those in the boundary mutation-drift Moran model, boundary terms are introduced to derive a system of inhomogeneous differential equations yielding the first two eigenvalues $\lambda_0=0$, and $\lambda_1=\theta$. Hence, the first non-zero eigenvalue is identical to that of the transition density of the general mutation model represented by orthogonal modified Jacobi polynomials (see Sec.~\ref{section:GeneralMutationDiffusion}). This representation of the transition density of the boundary mutation-drift Moran model can be used for inference with our forward-backward algorithm that combines the particle model and the orthogonal polynomial approach: For this to work, we must  establish a continuous prior distribution $\bs{\rho(x)}$ for the boundary mutation-drift Moran model (Sec.~\ref{sec:functional}). We will then review the forward-backward algorithm using orthogonal polynomials for the boundary mutation-drift Moran model (Sec.~\ref{section:for_back_gegenbauer}), and introduce the corresponding particle model (Sec.~\ref{section:particle_gegenbauer}).

\subsection{A functional as an improper prior} 
\label{sec:functional}

The beta equilibrium distribution Eq.~(\ref{eq:bb}) \citep{Wrig31} is generally assumed as the prior distribution $\phi(x\given t=S)=\bs{\rho}(x)$ for the bi-allelic Moran model in the continuous forward algorithms (Sec.~\ref{section:diffusion_Moran_for_back}). For the boundary mutation-drift Moran model, we define the functional
\begin{equation}\label{eq:Vogl_Bergman_functional}
 eq(x\given \alpha,\theta)=\lim_{N\to\infty}\begin{cases}
 \beta\big(1-\alpha\theta \int_{1/N}^{1-1/N} \frac1x\,dx\big) &\text{for $x=0$,}\\
 \alpha\beta\theta\,\frac{1}{x(1-x)} &\text{for $1/N \leq  x \leq 1-1/N$,}\\
 \alpha\big(1-\beta\theta \int_{1/N}^{1-1/N} \frac1x\,dx\big) &\text{for $x=1$,}
 \end{cases}
\end{equation}
to replace this beta distribution the prior. For the functional to be a valid prior, taking a sample of size $K$ with $y$ focal alleles with replacement from it should result in a marginal distribution for $y$ that is equal to the stationary distribution in Eq.~(\ref{eq:eq_boundary}). We will prove this here: 
\begin{itemize}
\item{(i)} For polymorphic samples $1\leq y\leq (K-1)$, the function 
\begin{equation}
\begin{cases} 
    \alpha\beta\theta\,x^{y-1}(1-x)^{K-y-1} &\text{within $[1/N,1-1/N]$}\\
    0 &\text{otherwise}
\end{cases}
\end{equation} 
is bounded from above by $\alpha\beta\theta\,x^{y-1}(1-x)^{K-y-1}$ within the interval $[0,1]$ (but converges monotonically towards it with increasing $N$). By the monotone convergence theorem, the order of taking the limit and integration can be reversed and we obtain:
\begin{equation}
\begin{split}
    \Pr(1\leq y\leq K-1\given K,\alpha,\theta)
    &=\alpha\beta\theta\binom{K}{y}\lim_{N\to\infty} \int_{1/N}^{1-1/N} x^{y-1}(1-x)^{K-y-1}\,dx\\
    &=\alpha\beta\theta \frac{K}{y(K-y)}\,.
\end{split}
\end{equation}
\item{(ii)} For monomorphic samples, analogous arguments apply. We show the case $y=K$ (equivalent calculations hold for $y=0$):
\begin{equation}
\begin{split}
    \Pr(y=K\given K,\alpha,\theta)&= \lim_{N\to\infty} \int_{1/N}^{1-1/N} x^K\alpha\beta\theta \frac{1}{x(1-x)}\\
    &\qquad+ \alpha\bigg(1-\beta\theta \int_{1/N}^{1-1/N} \frac1x\,dx\bigg) \,dx\\
    &= \lim_{N\to\infty} \int_{1/N}^{1-1/N} \alpha\beta\theta \frac{x^{K-1}}{1-x}+ \alpha\bigg(1-\beta\theta \int_{1/N}^{1-1/N} \frac1{1-x}\,dx\bigg) \,dx\\
   &= \alpha+\lim_{N\to\infty}\alpha\beta\theta \int_{1/N}^{1-1/N}  (x^{K-1}-1)(1+x+x^2+\dots) \,dx\\
   &=\alpha+\lim_{N\to\infty}\alpha\beta\theta \int_{1/N}^{1-1/N}  1+x+x^2+\dots+x^{K-2} \,dx\\
   &=\alpha-\alpha\beta\theta H_{K-1}\,.
\end{split}
\end{equation}
\end{itemize}

The functional Eq.~(\ref{eq:Vogl_Bergman_functional}) can thus be used as an improper prior distribution for the boundary mutation-drift Moran model. As long as $K<e^{\frac1{\max(\alpha,\beta)\theta}}$ approximately holds, a proper marginal posterior distribution for the proportion of focal alleles in the sample will result. 

\subsection{Transition probabilities with small scaled mutation rates}

\subsubsection{Forward-backward with augmented Gegenbauer polynomials}
\label{section:for_back_gegenbauer}

Recall that the forward and backward operators of the Kolmogorov diffusion equation for the general mutation bi-allelic Moran model can be decomposed into eigenvectors represented as (modified) Jacobi polynomials and corresponding eigenvalues  (Sec.~\ref{section:GeneralMutationDiffusion}, Sec.~\ref{section:diffusion_Moran_for_back}). A similar decomposition can be achieved for the boundary mutation-drift Moran model; we will reproduce an outline of the derivation here, details can be found in \citet[][Appendix]{Vogl16}: 

\paragraph{Pure drift model}
In a first step, the eigenvectors of a pure drift model are obtained by a zeroth order Taylor series expansion around $\theta$ in Sec.~(\ref{section:GeneralMutationDiffusion}), Sec.~(\ref{section:diffusion_Moran_for_back}).

The backward eigenvectors of the pure drift model are \citep{Berg17}: 
\begin{equation}\label{eq:backw_Us}
\begin{cases}
    B_0^{(\alpha,\theta)}(x)&=B_0^{(\alpha,0)}(x)=1\\
    B_1^{(\alpha,\theta)}(x)&=B_1^{(\alpha,0)}(x)=x-\alpha\\
    B_{n\geq2}^{(\alpha,0)}(x)&=w(x)U_n(x)\,,
\end{cases}
\end{equation}
where for $n\geq 2$:
\begin{equation}\label{eq:Gegenbauer_modified}
\begin{split}
  U_n(x)&=\sum_{l=0}^{n-2}(-1)^l\binom{n+l}{l}\binom{n-1}{l+1}x^l\,.
\end{split}
\end{equation}
are the modified Gegenbauer polynomials (compare \citep{Song12} Eqs.~(12),(13)) with weight function $w(x)=x(1-x)$. For any $m,n$, the modified Gegenbauer polynomials fulfill the following orthogonality relationship with respect to the weight function: 
\begin{equation}\label{eq:ortho_Gegen}
    \int_0^1 U_n(x) U_n(x) w(x)\,dx=\delta_{n,m} \Delta_n\,,
\end{equation}
where $\delta_{n,m}$ is Kronecker's delta and $\Delta_n=\frac{(n-1)}{(2n-1)n}$ for $n\geq 2$ are proportionality constants. The forward eigenvectors of the pure drift model are: 
\begin{equation}\label{eq:forw_Us}
\begin{cases}
    F_0^{(\alpha,0)}(x)&=\beta\delta(x)+\alpha\delta(x-1)\\
    F_1^{(\alpha,0)}(x)&=-\delta(x)+\delta(x-1)\\
    F_{n\geq2}^{(\alpha,0)}(x)&=-\frac{(-1)^n}n\delta(x)+U_n(x)-\frac{1}n\delta(x-1)\,.
\end{cases}
\end{equation}
The corresponding eigenvalues of the system are $\lambda_0=0$, $\lambda_1=\theta=0$, and $\lambda_n=n(n-1)$ for $n\geq 2$.

\paragraph{Boundary mutation-drift model}
To introduce mutations arising exclusively at the monomorphic boundaries, $\lambda_1$ is set to $\theta$ with $0<\theta<<1$. The discrete boundary mutation-drift model is scaled so that mutations enter the polymorphic region from the respective boundaries at average rates of $\frac{\alpha\theta}{N}$ and $\frac{\beta\theta}{N}$ per Moran drift event in equilibrium; this is required to maintain a constant equilibrium mutation rate across generations. In an orthogonal polynomial representation of the transition density of the continuous boundary mutation-drift model in which the (effective) population size and thus the generation lengths are not necessarily constant, mutations enter the polymorphic region from the respective boundaries at rates $\alpha\theta b_0(x,t)$ and $\beta\theta b_1(t)$: Here, $b_0(t)$ and $b_1(t)$ represent the probability masses both already fixed at the boundary at time $t$ or expected to fix there imminently by drift. In the polymorphic region, drift operates at an exponential rate that depends on the sample size $\kappa$, which is the order of the polynomial expansion and the sample size $K$. Overall, the forward and backward eigenvectors of the pure drift model of order $K$ are multiplied with the solution $\tau_n(t)$, with $0\leq n\leq K$, of the following system of first order linear equations describing the temporal dynamics induced by the boundary mutation (\citep{Vogl16}, but note the different weighting there):
\begin{equation}\label{eq:TauExpansion_origional}
 \begin{split}
   &\frac{d}{dt} \tau_0(t) = 0\\
   &\frac{d}{dt} \tau_1(t) = -\theta\tau_1(t)\\
   &\frac{d}{dt} \tau_n(t) = -\lambda_n\tau_n(t) + A_n +B_n e^{\theta t} \text{ for $n\geq 2$}\\
  \end{split} 
\end{equation}
where
\begin{equation*}\label{eq:TauExpansion_A}
    A_n = -\alpha\beta\theta(2n-1)n((-1)^n+1)
\end{equation*}
and
\begin{equation*}\label{eq:TauExpansion_B}
    B_n = -\theta(2n-1)n(b_0(t)-\beta)((-1)^n\alpha - \beta)\,.
\end{equation*}

\paragraph{Diagonalized boundary mutation-drift model} 
The system (Eq.~\ref{eq:TauExpansion_origional}) corresponds to a triangular matrix with eigenvalues $\lambda_0=0$, $\lambda_1=\theta$, $\lambda_n=n(n-1)$ on the main diagonal. Hence the eigensystem can be easily diagonalized \citep{Berg17}:

The backward eigenfunctions of the boundary mutation-drift model become:
\begin{equation}\label{eq:backw_bound_diag}
\begin{cases}
    B_0^{(\alpha,\theta)}(x)&=B_0^{(\alpha,0)}(x)=1\\
    B_1^{(\alpha,\theta)}(x)&=B_1^{(\alpha,0)}(x)=x-\alpha\\
    B_{n \geq 2}^{(\alpha,\theta)}(x)&=B_n^{(\alpha,0)}(x)-\vartheta \frac{E_n\Delta_n}{\lambda_n} B_0^{(\alpha,0)}(x)-\theta \frac{B_n\Delta_n}{\lambda_n} B_1^{(\alpha,0)}(x)\,.
\end{cases}
\end{equation}
with
\begin{equation}
\begin{split}
    \vartheta&=\alpha\beta\theta\,,\\
    E_{n}&=-(n-1)\frac{((-1)^n+1)}{\Delta_n}\,,\\
    O_{n}&=-(n-1)\frac{(-1)^n\alpha-\beta}{\Delta_n}\,,
\end{split}
\end{equation}
Equivalently, the forward eigensystem becomes \citep{Berg17}: 
\begin{equation}\label{eq:forw_bound_diag}
\begin{cases}
    F_0^{(\alpha,\theta)}(x)&=eq(x\given \alpha,\theta)\\
    F_1^{(\alpha,\theta)}(x)&=neq(x\given \alpha,\theta)\\
    F_{n \geq 2}^{(\alpha,\theta)}(x)&=F_n^{(\alpha,0)}(x)\,,
\end{cases}
\end{equation}
with  the functional $eq(x\given \alpha,\theta)$ from Eq.~(\ref{eq:Vogl_Bergman_functional}), and the similar functional: 
\begin{equation}\label{eq:Vogl_Bergman_functional_1}
 neq(x\given \alpha,\theta)=\lim_{N\to\infty}\begin{cases}
 -1+\alpha\theta \int_{1/N}^{1-1/N} \frac1x\,dx &\text{for $x=0$,}\\
 \theta\,\bigg(-\frac{\alpha}{x} + \frac{\beta}{1-x}\bigg)&\text{for $1/N \leq  x \leq 1-1/N$,}\\
 1-\beta\theta \int_{1/N}^{1-1/N} \frac1x\,dx &\text{for $x=1$.}
 \end{cases}
\end{equation}
The weights are augmented by $\Delta_0=\Delta_1=1$.

Note that the first diagonalized eigenfunction $F_0^{(\alpha,\theta)}(x)$ corresponds to the equilibrium distribution. With this and the second eigenfunction $F_1^{(\alpha,\theta)}(x)$ a quasi-equilibrium state of the allele proportions in the polymorphic interior can be modelled:
\begin{equation}
    F_0^{(\alpha,\theta)}(x)+\alpha e^{-\theta t} F_1^{(\alpha,\theta)}(x)\,.
\end{equation}
Observe that the rate of approach to equilibrium depends on the mutation rate $\theta$, which is considerably slower than the rate of drift. 


\paragraph{Application of the forward-backward algorithm} 

If the ancestral allele configuration \bs{\rho}(x) is taken to be distributed  according to the functional in Sec.~(\ref{sec:functional}) with
\begin{equation}
    \bs{\rho}(x)=\sum_{n=0}^\infty \rho_n^{(\alpha,\theta)}   F_{n}^{(\alpha,\theta)}(x),
\end{equation}
and $\rho_n^{(\alpha,\theta)}$ obtained according to:
\begin{equation}\label{eq:forward_given_rho_coefs_2}
    \rho_n^{(\alpha,\theta)}= \frac{1}{\Delta_n^{(\alpha,\theta)}} \int_{x=0}^1 w(x)  F_n^{(\alpha,\theta)}(x)\bs{\rho}(x)dx\,.
\end{equation}

The forward transition density of the population can be expanded to either: 
\begin{equation}\label{eq:forward_given_rho_geg}
\phi(x\given t,\bs{\rho})=w(x) \sum_{n=0}^\infty \rho_n^{(\alpha,\theta)} F_n^{(\alpha,0)}(x) \tau_n(t)\,,
\end{equation}
or
\begin{equation}\label{eq:forward_given_rho_geg_diag}
\phi(x\given t,\bs{\rho})=w(x) \sum_{n=0}^\infty \rho_n^{(\alpha,\theta)} F_n^{(\alpha,\theta)}(x) (\tau_n(t))^-\,,
\end{equation}
where $(\tau_n(t))^-$ is the solution to 
\begin{equation}\label{eq:TauExpansion_2}
 \begin{split}
   &\frac{d}{dt} \tau_n(t) = -\lambda_n\tau_n(t)
  \end{split} 
\end{equation}

The marginal likelihood of the number of focal alleles $y$ in an extant sample of size $K$ can similarly be determined via:
\begin{equation}\label{eq:marg_like_general_geg}
\begin{split}
\Pr(\y \given K,\alpha,\theta,t=0,\bs{\rho})=\sum_{n=0}^K \rho_n^{(\alpha,\theta)} d_n(K,y) \Delta_n \tau_n(t=0)\,. 
\end{split}
\end{equation}
or
\begin{equation}\label{eq:marg_like_general_geg_diag}
\begin{split}
\Pr(\y \given K,\alpha,\theta,t=0,\bs{\rho})&=\sum_{n=0}^K \rho_n^{(\alpha,\theta)} d_n(K,y) \Delta_n  (\tau_n(t=0))^-\,, 
\end{split}
\end{equation}

with

\begin{equation}
    d_n(K,y)=\frac{1}{\Delta_n}\int_{x=0}^1 \binom{K}{y}x^y(1-x)^{K-y}\,F_n^{(\alpha,0)}(x)\,dx\,.
\end{equation}

\paragraph{Summary} A forward-backward algorithm approach using orthogonal polynomials can easily be applied to the boundary mutation-drift Moran model. Note that using the un-diagonalized version of the eigensystem is particularly efficient when mutation parameters change over time, both with respect to time and numerical accuracy, since this change does not affect the eigenvectors but only the temporal system. Although the boundary mutation-drift Moran model is an approximation to the general mutation bi-allelic Moran model, it may provide more accurate results in an inference framework than direct use of the general mutation model through this numerical advantage. The increased numerical accuracy may allow the use of higher sample sizes, which is necessary for many data analyses.

\subsubsection{Particle transition probabilities}
\label{section:particle_gegenbauer}

As with the general mutation model, Eq.~(\ref{eq:marg_like_general_geg}) and Eq.~(\ref{eq:marg_like_general_geg_diag}) can be determined with an appropriate particle model. This can again be constructed using the ratios of the probabilities of the appropriate configurations of ancestral focal alleles $\upsilon$ in a sample size of $\kappa$ (see Eq.~\ref{eq:ratio_general}).

Recall that the indicator variable $z_{\kappa+1}$ is one if the $\kappa+1$st allele is of the focal type and zero otherwise.
Note that the transition probabilities from $z_{\kappa}$ to $z_{\kappa+1}$ must hold in equilibrium. Recall that for the general mutation model, we have:
\begin{equation}
\begin{split}
 {\Pr}(z_{\kappa+1}\given \upsilon,\kappa,\alpha,\theta)&=\frac{{\Pr}(z_{\kappa+1}+\upsilon\given \kappa+1,\alpha,\theta)}{{\Pr}(\upsilon\given \kappa,\alpha,\theta)}\\
&=\frac{(\upsilon+\alpha\theta)^{1-z_\kappa}(\kappa-\upsilon+\beta\theta)^{z_\kappa}}{\kappa+\theta}\,.
\end{split}
\end{equation}
Note that the terms of  ${\Pr}(z_{\kappa+1}+\upsilon\given \kappa+1,\alpha,\theta)$ and ${\Pr}(\upsilon\given \kappa,\alpha,\theta)$ that only hold in equilibrium cancel out.  Therefore, the transition probabilities for the general mutation model as given in Eq.~(\ref{eq:ratio_general}) hold in non-equilibrium situations. 

For the boundary mutation-drift model, transition probabilities need to be consistent with the equilibrium probability Eq.~(\ref{eq:Vogl_Bergman_functional}). For $1\leq \upsilon\leq (\kappa-1)$, no mutations are possible and the transition probabilities are independent of the mutation parameters $\alpha$ and $\theta$:
\begin{equation}\label{eq:trans_poly}
\begin{split}
    {\Pr}(z_{\kappa+1}\given \upsilon,\alpha,\theta)&=\frac{{\Pr}(z_{\kappa+1}+\upsilon\given \kappa+1,\alpha,\theta)}{{\Pr}(\upsilon\given \kappa,\alpha,\theta)}\\
    &=\frac{(\kappa-\upsilon)^{z_\kappa} \upsilon^{1-z_\kappa}} {\kappa}\,.
\end{split}
\end{equation}
For $\upsilon=0$, the transition probability consistent with that in Eq.~(\ref{eq:trans_poly}) is:
\begin{equation}
\begin{split}
    {\Pr}(z_{\kappa+1}\given \upsilon=0,\alpha,\theta)&=\frac{{\Pr}(z_{\kappa+1}\given \kappa+1,\alpha,\theta)}{{\Pr}(\upsilon=0\given \kappa,\alpha,\theta)}\\
    &=\frac{(\frac{\alpha\theta}\kappa)^{z_\kappa}(1-\alpha\theta H_{\kappa})^{1-z_\kappa}} {1-\alpha\theta H_{\kappa-1}}\,,
\end{split}
\end{equation}
and the analogous transition rate holds for $\upsilon=\kappa$. 

For the transition from $\kappa=1$ to $\kappa=0$ (or the reverse forward in time), the transition probabilities of the boundary mutation-drift models are identical to those of the general mutation model: Specifically, the probabilities are $\alpha$ and $\beta$ to zeroth order in $\theta$. 

\paragraph{Temporal dynamics}
 
 The backward temporal dynamics of the particle model are clearly equivalent to those in Sec.~(\ref{section:backward-particle_temp}) however with eigenvalues $\lambda_0=0$, $\lambda_1=\theta$, and
 $\lambda_n=n(n-1)$ for $n\geq2$.

\paragraph{Summary}
A particle model can be used to trace sample allele configurations in the boundary mutation-drift Moran model forward and backward in time. The transition probabilities defined through it are sample consistent, and the first non-zero eigenvalue (defining the speed of approach to equilibrium) is identical to that of the general mutation Moran model and independent of the sample size. In phylogenetic settings, where split times (in generations) between species or populations are large compared to effective population sizes, this eigenvalue dominates. 

\section{Conclusions}

In population genetics, small to moderately sized extant samples are generally used to infer parameters of past population processes. The evolutionary trajectory of the population can, under certain assumptions, be described by the bi-allelic decoupled Moran model in the diffusion limit.  

To determine the marginal likelihood of sample allele proportions in the extant sample, the backwards N-particle diffusion of the extant sampling distribution can be decomposed into eigenvectors of orthogonal polynomials and corresponding eigenvectors. The backward orthogonal polynomial approach is computationally efficient and accommodates temporally changing effective population sizes (and therefore mutation rates) or mutation biases. This is even more true when scaled overall mutation rates are small and a boundary mutation-drift Moran model is assumed to describe the evolutionary dynamics: if the system is not diagonalized, time inhomogeneous dynamics allow for effient numerics. 

The probabilities of ancestral sample allele configurations can be inferred with a so-called particle model augmented with backwards temporal dynamics: It directly traces the size and allele proportions of a sample both forward (urn model) and backward (coalescent) in time. Empirically, this approach may be useful for inferring the timing of mutation-drift events at a specific locus. However, its main value lies in its equivalence with the backwards orthogonal polynomial method: The backward orthogonal polynomial expansion of an extant sample to $n$th order has the genealogy of the sample embedded within it. It therefore contains all the information traditionally modelled by the coalescent, but is more tractable and numerically superior. 

In order to infer the probability of ancestral population allele proportions at any time in the past, the particle model needs to be complemented with a forward in time orthogonal polynomial approach. The result is a full forward-backward algorithm equivalent to the forward-backward algorithm using only orthogonal polynomials.

\section*{Acknowledgments}

We thank the present and past members of the doctorate college population genetics, especially Juraj Bergman, for stimulating discussions. CV and SP's research was supported by the Austrian Science Fund (FWF): DK W1225-B20; LCM's by the School of Biology at the University of St.Andrews and also partially funded through Vienna Science and Technology Fund (WWTF) [MA016-061].

\bibliography{coal}

\end{document}